\documentclass[prb,twocolumn,floatfix,showpacs]{revtex4}
\usepackage{graphicx}
\usepackage{amssymb}
\usepackage{amsmath}

\begin{document}
\title{Charging of a quantum dot coupled to Luttinger liquid leads} 
\author{P.~W\"achter} \affiliation{Institut
  f\"ur Theoretische Physik, Universit\"at G\"ottingen, 37077
  G\"ottingen, Germany} 
\author{V.~Meden} \affiliation{Institut f\"ur
  Theoretische Physik, Universit\"at G\"ottingen, 37077 G\"ottingen,
  Germany} 
\author{K.~Sch\"onhammer} \affiliation{Institut f\"ur
  Theoretische Physik, Universit\"at G\"ottingen, 37077 G\"ottingen,
  Germany}

\date{\today}

\begin{abstract}
Luttinger liquid behavior of one-dimensional correlated electron
systems is characterized by power-law scaling of a variety of physical 
observables with exponents determined by a single interaction
dependent parameter $K$. We suggest a setup to study Luttinger 
liquid behavior in quantum wires which allows to determine $K$ from
two independent measurements: transport through a quantum dot
embedded in the wire and the charge on the dot. Consistency of the
two $K$'s for a single probe would provide strong
experimental evidence for the Luttinger liquid paradigm. 
\end{abstract}
\pacs{71.10.Pm, 73.23.Hk, 73.40.Gk}
\maketitle     

\section{Introduction}

Theoretically it is well 
established that the two-particle interaction $U$
in metallic, one-dimensional (1d) electron systems leads to Luttinger
liquid (LL) physics.\cite{Schoenhammer05} One of the characterizing
properties is the power-law scaling of a variety of physical observables
as functions of external parameters (e.g.~the temperature $T$)
with exponents which can be expressed in terms of a {\it single,}
interaction dependent LL parameter $K \leq 1$ (for repulsive
interactions, with $K=1$ for $U=0$). However, there are only a few
experiments on quasi 1d systems which reveal clear indications 
of LL behavior. Even in these rare examples mostly only a {\it single}
observable as a function of a {\it single} control parameter was measured. 
In this case it is very difficult to convincingly exclude any 
other source for the observed power-law scaling 
than LL physics.  
The situation becomes more complex as even for the
same type of quantum wire (e.g. semi-conductor heterostructures,
carbon nanotube, or a row of atoms on a surface) $K$ might vary 
from probe to probe since it depends not only on $U$ but also on other 
details such as the band structure and filling.\cite{Schoenhammer05} 

A more direct evidence for LL behavior could be achieved in the following
way. Using a {\it single probe} one should measure {\it two observables} for
which LL theory predicts power-law scaling with different exponents
$\beta_1(K)$ and $\beta_2(K)$ -- in the optimal case even as 
functions of {\it two different control
parameters.} If the two exponents turn out to be consistent, that is
$K(\beta_1) \approx K(\beta_2)$, strong evidence for LL physics is
achieved. A step in this direction is the linear conductance $G(T)$
measurement by Yao {\it et al.}\cite{Yao} across an impurity free part of a
metallic single-wall carbon nanotube as well as across 
a part of the {\it same}
tube containing a single kink (impurity). 

\begin{figure}[t]
\begin{center}
\includegraphics[width=0.35\textwidth,clip]{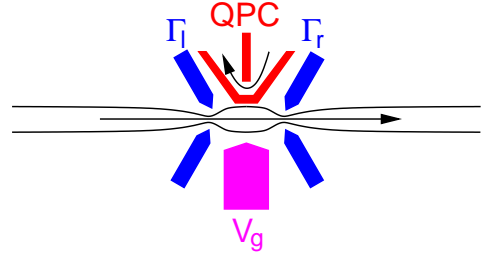}
\end{center}
\vspace{-0.6cm}
\caption[]{(Color online) Sketch of the setup to measure
  transport through a quantum dot as well as the charge on the dot. 
\label{skizz}}
\end{figure}

We propose the setup sketched in
Fig.~\ref{skizz}, in which one can measure $G$ 
through a quantum dot (QD) embedded in a 1d wire as a function 
of $T$ and, for the same probe, the charge $n$
accumulated on the dot as a function of the dot level position
varied by an external gate voltage $V_g$. 
The QD is formed by two high barriers within the 1d system, e.g.~realized
by additional gates. The charge is detected by the current running through
a nearby quantum point contact (QPC). While transport of 1d
correlated electrons through double barriers has extensively been
studied theoretically in recent years,\cite{doublebarrier} 
less is known about the charging of a {\it small} QD coupled to 
two LL leads.\cite{footnote} 
Here we investigate in detail how $n(V_g)$ is affected by 
LL physics using two approaches. First we consider a 
field-theoretical, effective low-energy model, the infinite Luttinger 
model,\cite{Schoenhammer05} 
and perturbation theory in the dot-LL coupling $\Gamma$. This 
can be done for arbitrary $0 < K \leq  1$, but is restricted to 
small $\Gamma$. In a complementary, second step we study an
interacting 
microscopic lattice model of finite length coupled to noninteracting 
leads, a model being closer to experimental setups.
To treat the correlations we use the functional renormalization 
group (fRG).\cite{fRGbasic} This method can be applied for arbitrary 
$\Gamma$, but is restricted to small $1-K$. 
Both approaches lead to consistent results and we show that 
$n(V_g)$ is governed by power-law scaling, which should be detectable 
in the suggested setup. 
We consider a dot with a large level spacing such
that only a single level matters. Furthermore, we mainly consider 
spinless fermions and suppress the Kondo effect. Experimentally this can
be achieved by a magnetic field lifting the spin degeneracy of the
dot level or by measuring at $T > T_K$, with $T_K$ being the Kondo
temperature. For transport through a dot showing the Kondo effect 
coupled to LL leads, see Ref.~\onlinecite{STV}. 

From the linear conductance $G(V_g,T)$ through a double barrier 
$K$ can be extracted in several ways,   
some of them restricted to certain regimes of $K$ values or 
symmetric barriers.\cite{doublebarrier} 
To be as general as possible we here 
present a prediction which holds for all $0<K<1$, symmetric as well
as asymmetric barriers, and which does not require any other fine tuning 
of parameters. For a fixed gate voltage away from 
resonance, which we assume to be at $V_g=0$, one finds 
$G \sim \max\{T,\delta\}^{2 (1/K-1)}$ at asymptotically small 
scales.\cite{doublebarrier,footnoteresowidth} 
Here $\delta$ denotes an energy scale $\sim 1/N$, with $N$ 
being the length of the LL wire, which is
eventually coupled to noninteracting leads. 

In an important work Furusaki and Matveev analyzed $n(V_g)$ for 
strongly interacting systems with $K<1/2$ within the infinite Luttinger 
model using perturbation theory in $\Gamma$ and the mapping 
to related problems.\cite{Furusaki} They showed that for sufficiently 
small $\Gamma$, $n(V_g)$ is discontinuous at $V_g=0$. For $1/3 < K 
< 1/2$ the finite $V_g$ behavior adjacent to the jump shows scaling 
with the exponent $1/K-2$, while for even smaller $K$ the deviations
are linear in $V_g$. The perturbation theory in $\Gamma$ 
for the Green function -- not for the self-energy, as used by us -- breaks 
down for $1/2< K <1$. In an attempt to investigate LLs characterized
by such $K$'s a numerical method was used for systems of up to 150 sites 
in Ref.~\onlinecite{Berkovits}. The authors concluded that $n(V_g)$ is 
continuous and does not show LL physics. 
Below we confirm the first statement but show that the second 
is incorrect as  finite size corrections completely mask the 
power-law behavior. 

\section{Perturbation theory in the level-lead coupling 
for the semi-infinite Luttinger model}

We first 
consider a QD coupled to two LLs via tunnel barriers with 
hopping amplitudes $t_{l/r}$. For simplicity the  LLs are assumed to be 
equal and described by the semi-infinite Luttinger 
model \cite{Schoenhammer05} (with an open 
boundary on the side coupled to the dot). To 
leading order in $\Gamma=t_l^2+t_r^2$ the dot self-energy 
is given by $\Sigma_d(z) = \Gamma {\mathcal G}(z)$, with 
the single-particle Green function $\mathcal G$ of the 
disconnected semi-infinite LL at the boundary. 
The low-energy behavior of the imaginary 
part  of ${\mathcal G}$ for $z=\omega+i0$, that is the spectral function 
$\rho$, is known exactly from 
bosonization.\cite{KaneFisher} It is given by $\rho(\omega) \sim 
|\omega|^{1/K-1}$. To be specific we assume that $ \rho(\omega) $  
has support $[-\omega_c,\omega_c]$ 
\begin{eqnarray}
\label{specLL}
\omega_c \, \rho(\omega) = \theta(\omega_c - |\omega|) \, 
|\omega/\omega_c|^{1/K-1}/(2K) \; . 
\end{eqnarray} 
It is then straight forward to compute 
$\mbox{Re} \,{\mathcal G}^R(\omega)$ 
by Hilbert transformation. 
The leading behavior at $|\omega/\omega_c| \ll 1$ 
is given by  
\begin{eqnarray}
\label{realpart}
\omega_c \, \mbox{Re} 
\,{\mathcal G}^R(\omega) \sim 
\left\{ \begin{array}{ll}
- \mbox{sign}\,(\omega) \, \left| \frac{\omega}{\omega_c}\right|^{1/K-1} 
& \mbox{for} \, \frac{1}{2}< K<1 \\
 \frac{\omega}{\omega_c}\, \ln{ \left| \frac{\omega}{\omega_c}\right|}
& \mbox{for} \, K=\frac{1}{2} \\ 
-  \frac{\omega}{\omega_c}
& \mbox{for} \, K<\frac{1}{2} \; .
\end{array}\right.
\end{eqnarray}
Using the Dyson equation the dot spectral function $\rho_d$ follows 
from the perturbative $\Sigma_d$ as 
\begin{eqnarray}
\label{dotspectral}
\rho_d(\omega) = \frac{\Gamma \rho(\omega)}{\left[ \omega -V_g - \Gamma 
 \mbox{Re} \,{\mathcal G}^R(\omega)\right]^2+ \left[\pi \Gamma 
\rho(\omega) \right]^2} \; .
\end{eqnarray}
The dot charge is
\begin{eqnarray}
\label{charge}
n(V_g) = \int_{- \omega_c}^\mu d \omega \; \rho_d(\omega) \; ,
\end{eqnarray}
with the chemical potential $\mu=0$. Because of the 
particle-hole symmetry it obeys $n(V_g) = 1- n(-V_g)$ and from now on
we focus on $V_g \geq 0$. In contrast 
to the perturbation theory in $\Gamma$ for the dot Green function
itself used 
in Ref.~\onlinecite{Furusaki} which is restricted to $K<1/2$, our approach 
can be applied for all $0<K \leq 1$.  

Based on Eqs.~(\ref{specLL})-(\ref{charge}) the leading small $V_g$
behavior of $1/2-n(V_g)$ can be determined analytically. 
For $1/2 < K \leq 1$, $n(V_g)$ is a continuous function with 
$n(V_g=0)=1/2$. This implies that the width $w$ over which $n(V_g)$
changes from $1$ to $0$ is finite.\cite{footnoteresowidth}
The function $n(V_g)$ contains regular terms proportional 
to $V_g^{2l+1}$, with 
$l \in \mathbb{N}_0$, as well as anomalous terms with exponents 
containing $K$. The leading anomalous term is $\sim
V_g^{(2K-1)/(1-K)}$. Depending on $K$ either the linear term or the
anomalous term dominates. A special situation is reached at $K=2/3$,
where logarithmic corrections appear. The leading $V_g$ dependence is
given by 
\begin{eqnarray}
\label{leadingn}
\frac{1}{2}-n(V_g) \sim 
\left\{ \begin{array}{ll}
\frac{V_g}{\omega_c} & \mbox{for} \, \frac{2}{3}< K \leq 1 \\
\frac{V_g}{\omega_c} \, 
\left|\ln{ \left( \frac{V_g}{\omega_c}\right)} \right|
& \mbox{for} \, K=\frac{2}{3} \\ 
\left( \frac{V_g}{\omega_c}\right)^{(2K-1)/(1-K)}
& \mbox{for} \, \frac{1}{2} < K<\frac{2}{3} \; .
\end{array}\right.
\end{eqnarray}  

At $K=1/2$, $n(V_g)$ is still continuous and for $V_g \searrow 0$ 
approaches $1/2$ with corrections $\sim 1/|\ln(V_g/\omega_c)|$. 

For $K<1/2$ and small $\Gamma$, $n(V_g)$ shows a jump at 
$V_g=0$, that is $\lim_{V_g  \searrow 0} n(V_g) = \Delta < 1/2$. 
In this regime our perturbative approach for the self-energy, 
which guarantees the correct analytical structure of the dot Green 
function, gives the same results as the perturbation theory for the 
Green function itself used in Ref.~\onlinecite{Furusaki}. This follows from two 
observations. According to Eq.~(\ref{realpart}) the real part of 
${\mathcal G}^R$ becomes linear at small $\omega$ and can thus 
be absorbed in the first term in the denominator of 
Eq.~(\ref{dotspectral}). In addition, for small $V_g$ the 
contribution of $\rho$ in the denominator of 
Eq.~(\ref{dotspectral}) can be neglected compared to the term 
linear in $\omega$. For the small $V_g$ analysis and to leading order 
in $\Gamma$ the integrand in Eq.~(\ref{charge}) becomes equivalent
to the one obtained in Ref.~\onlinecite{Furusaki}
\begin{eqnarray}
\label{chargesimp}
n(V_g) \sim \Gamma \int_{0}^{\omega_c} d \omega \, 
\frac{\omega^{1/K-1}}{(\omega+V_g)^2} \;.
\end{eqnarray}
The jump at $V_g \searrow 0$ is given by 
$ \Delta = \Gamma /[(2-4K)\omega_c^2]$ 
which is nonuniversal as it depends on the cutoff $\omega_c$. 
Evidently, for $K$ close to $1/2$ this expression only holds for
sufficiently small $\Gamma$.
In Ref.~\onlinecite{Furusaki} it is argued that increasing $\Gamma$
beyond the perturbative regime $\Delta$ decreases, approaches 
the minimal value $\Delta_{0}=\sqrt{K/2}$ at a certain
$\Gamma_{0}$, and  
for $\Gamma > \Gamma_{0}$, $n$
becomes a continuous function of $V_g$ even for $K<1/2$. 
The finite $V_g$ corrections of $n$ for small $\Gamma$ 
are given by 
\begin{eqnarray}
\label{leadingdelta}
\Delta-n(V_g) \sim 
\left\{ \begin{array}{ll}
\left( \frac{V_g}{\omega_c} \right)^{1/K-2}& 
\mbox{for} \, \frac{1}{3}< K< \frac{1}{2} \\
\frac{V_g}{\omega_c} \, \left|\ln{ \left( \frac{V_g}{\omega_c}\right)} \right|
& \mbox{for} \, K=\frac{1}{3} \\ 
\frac{V_g}{\omega_c}
& \mbox{for} \, 0< K<\frac{1}{3} \; .
\end{array}\right.
\end{eqnarray} 

These results show that for $1/3< K< 2/3$, that is for sufficiently 
strong, but not too strong interactions, the LL parameter $K$ can be 
extracted from a measurement of $n(V_g)$ for gate voltages 
close to the resonance value. 

A second way to extract the LL parameter in the regime in which $n(V_g)$ 
is continuous, that is for $1/2<K<1$, is given by the $\Gamma$ dependence
of the characteristic width $w$ over which the charge changes from $1$
to $0$. In particular, this includes weak interactions with $2/3 <K<1$ 
for which $1/2-n(V_g)$ itself is linear in $V_g$ and cannot directly
be used to determine $K$. The width can e.g.~be defined by $w=2
V_g^{0}$ with $n(V_g^{0}) \equiv 1/4$. In experimental setups in
which the two barriers are realized by gates, $\Gamma$ can be tuned by
varying the applied voltages and $w(\Gamma)$ can be extracted. For 
$\Gamma \to 0$, $w(\Gamma)$ follows from Eq.~(\ref{chargesimp}) and 
scales as 
\begin{eqnarray}
\label{wid} 
\frac{w(\Gamma)}{\omega_c} \sim \left( \frac{\Gamma}{\omega_c^2}
\right)^{K/(2K-1)} \;\;\;\;\;\;  \mbox{for} \; 1/2<K \leq 1 \; .
\end{eqnarray} 
On first glance the appearance of an anomalous exponent in $w$ 
might be at odds with the linear $V_g$ dependence of $1/2-n(V_g)$ for
$2/3 < K <1$.  In fact, both results are consistent as the regime 
over which $n(V_g)$ goes linearly through $1/2$ around $V_g \approx 0$ 
shrinks with decreasing $\Gamma$ and decreasing $K$. To experimentally
observe the predicted power-law scaling the temperature has to be
sufficiently smaller than the width $w$. 

In the absence of the Kondo effect (see above) including 
the spin degree of freedom does not lead to new physics. The
perturbative analysis can be repeated after replacing the exponent
$1/K-1$ in the spectral function of Eq.~(\ref{specLL}) by the exponent 
for LLs with spin $(1/K-1)/2$.\cite{Schoenhammer05} 

\section{Weak to intermediate interactions in a microscopic 
lattice model}

We next replace the LLs described 
by the semi-infinite Luttinger model
by the microscopic lattice model with nearest-neighbor hopping $t>0$ and
nearest-neighbor interaction $U$. On both sides of the QD the LLs 
are assumed to be finite, each having $ \approx N/2$ sites, and 
adiabatically coupled to noninteracting 1d tight-binding 
leads.\cite{fRGbasic} The interaction
is treated by an approximation scheme that is based on the fRG and
which was shown to be reliable for weak to intermediate 
interactions.\cite{fRGbasic} In contrast to the perturbation theory in 
$\Gamma$, which is restricted to small $\Gamma$, this method can be 
applied for all $\Gamma$ and is thus complementary to the above
approach. The Hamiltonian is given by      
\begin{eqnarray}
\label{ham}
H & = & - t \sum_{j=-\infty}^{\infty}\left(c^\dagger_{j+1}c_j+
\textnormal{H.c.}\right) + V_g n_{j_d}\nonumber \\*
& + & \sideset{}{'} \sum_{j=1}^{N-1} U_{j,j+1}\left(n_j-\frac{1}{2}\right)
\left(n_{j+1}-\frac{1}{2}\right) \nonumber \\*
& - & (t_l -t) c^\dagger_{j_d}c_{j_d-1}^{} - (t_r -t ) 
c^\dagger_{j_d+1}c_{j_d}^{}-\textnormal{H.c.}  
\end{eqnarray}
in standard second quantized notation, with $n_j = c_j^\dag c_j$.  
To prevent any backscattering from the contacts to the noninteracting
leads around $j \approx 1$
and $N$ the interaction is turned on and off smoothly over a few
lattice sites, with a bulk value $U$, as described in 
Ref.~\onlinecite{fRGbasic}. The dot is located at lattice site $j_d$
somewhere close to $N/2$ (the results are insensitive to the exact
position). The prime at the sum in the second line
indicates, that the interaction across the barriers defining the QD 
is set to zero. 
We also studied the case in which the interaction on these bonds 
takes the bulk value $U$ and found that our conclusions 
are valid also for this
setup. We focus on half-filling of the band. In this
case the bulk model is a LL for $|U|< 2t$ and a closed expression 
for $K$ in terms of the model parameters can be given \cite{Haldane80}
$K^{-1}=2\arccos\left[-U/(2t)\right]/\pi$.

\begin{figure}[t]
\begin{center}
\includegraphics[width=0.45\textwidth,clip]{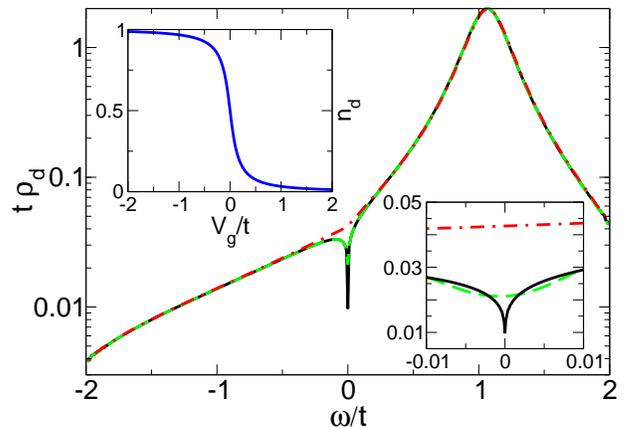}
\end{center}
\vspace{-0.6cm}
\caption[]{(Color online) Main part: Functional RG results for the 
  dot spectral function as a  function of energy for $U/t=0.5$, 
  $t_l/t=t_r/t=\sqrt{0.1}$, $V_g/t=1$, and
  length of the LL wire $N=10^2$ (dashed dotted line), $N=10^4$ 
  (dashed line), $N=10^5$ (solid line). Note the log-scale of the
  $y$-axis. Lower inset: Zoom-in of $\rho_d$ around $\omega=0$ 
  (linear-scale of $y$-axis). Upper inset: Dot occupancy as a function
  of the gate voltage for the same parameters as in the main part and
  $N=10^5$.   
\label{fig2}}
\end{figure}

Within the fRG one introduces an energy cutoff $\Lambda$ into the
noninteracting propagator. Taking the derivative of the generating
functional of the one-particle irreducible vertices with respect 
to $\Lambda$ and neglecting higher order
corrections one derives a set of ${\mathcal O}(N)$ 
coupled differential  equations for the $\Lambda$-flow of 
the self-energy and a renormalized nearest-neighbor interaction. 
It can be solved numerically for up to $10^7$ sites, 
resulting in an approximate expression for the dot Green 
function. This approach is described in detail in Ref.~\onlinecite{fRGbasic}.

From the Green function the spectral function $\rho_d$ and thus  
the charge on the dot [see Eq.~(\ref{charge})] can be computed. 
In Fig.~\ref{fig2} $\rho_d(\omega)$ is shown for $U=0.5$, symmetric
barriers $t_l/t=t_r/t=\sqrt{0.1}$, $V_g/t=1$, and different $N$ 
(note the log-scale of the $y$-axis in the main part). The upper 
inset shows $n(V_g)$ for $N=10^5$. On the scale 
of the plot $n(V_g)$ does not change if one further increases $N$.
The dominating feature of $\rho_d$ is the Lorentzian-like peak at $\omega
\approx V_g$.  Although a fermion occupying the dot is assumed 
to be noninteracting with the fermions in the leads, increasing $N$ the
coupling to the LL wires clearly leads to a power-law suppression 
$\rho_d(\omega) \sim \omega^{1/K-1}$ close to $\omega=0$, as also 
given by the perturbative expression Eq.~(\ref{dotspectral}). The
lower inset of Fig.~\ref{fig2} shows a zoom-in of the dip region.  
The absence of this LL feature at small $N$ of order $100$ explains
why in Ref.~\onlinecite{Berkovits} it was possible to fit $n(V_g)$ by a
Fermi liquid form. 

\begin{figure}[t]
\begin{center}
\includegraphics[width=0.45\textwidth,clip]{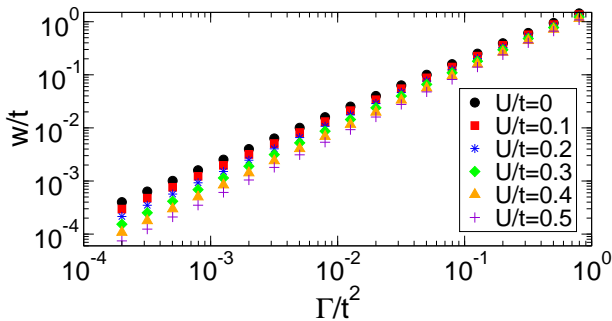}
\end{center}
\vspace{-0.6cm}
\caption[]{(Color online) Functional RG results 
   for the width $w$ \cite{footnotewdef} over which $n(V_g)$
  changes from 1 to 0 as a function of the dot-LL coupling 
  $\Gamma$ for $N=10^5$ and different $U$.     
\label{fig3}}
\end{figure}

\begin{figure}[t]
\begin{center}
\includegraphics[width=0.45\textwidth,clip]{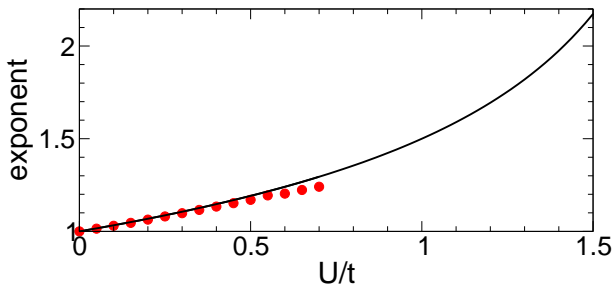}
\end{center}
\vspace{-0.6cm}
\caption[]{(Color online) Circles: Scaling exponent of the 
  width $w(\Gamma)$ over which $n(V_g)$ changes from 1 to 0 
  extracted from the fRG data for the microscopic model. 
  Solid line: The exponent $K/(2K-1)$ [see Eq.~(\ref{wid})], with
  $K=K(U)$, as obtained within the Luttinger model.   
\label{fig4}}
\end{figure}

The LL suppression of $\rho_d$ around $\omega = 0$ will manifest
itself also in the charging of the dot. To illustrate this we  
confirm the prediction of Eq.~(\ref{wid}) for the $\Gamma$ dependence of
$w$. We extract $w$ \cite{footnotewdef} 
from $n(V_g)$ (for an example of $n(V_g)$ 
see the inset of Fig.~\ref{fig2}) for $N=10^5$, a variety of $\Gamma$ (for
simplicity assuming symmetric barriers), and 
different $U$. The results for $w(\Gamma)$ are shown in 
Fig.~\ref{fig3} on a log-log scale. At small $\Gamma$, 
$w$ shows power-law
scaling. In Fig.~\ref{fig4} the exponent as a function of $U$, 
obtained by fitting the data of Fig.~\ref{fig3} 
(and additional data sets), is compared to $K/(2K-1)$ determined 
in perturbation theory in $\Gamma$ [see Eq.~(\ref{wid})]. 
We used the exact relation between $K$ and $U$ mentioned 
above. The results agree quite well for $0 \leq U/t \lesssim 1/2$. For 
larger $U$ higher order corrections neglected in our truncated fRG 
scheme become important. For sufficiently large $U$ the exponent 
$K/(2K-1)$ becomes large (it diverges for $K \searrow 1/2$) and 
should experimentally be clearly distinguishable from the 
noninteracting value $1$.

\section{Summary}

Using two different models and methods we have
investigated the charge $n(V_g)$ accumulated on a QD coupled to two 
LL wires when the dot level position is varied by an external 
gate voltage. Depending on the strength of the two-particle
interaction $U$, LL physics manifests itself in power-law scaling 
of $n(V_g)$ close to the resonance at $V_g=0$ 
and the width $w(\Gamma)$ over which $n(V_g)$ changes from $1$ to $0$.
The corresponding exponents can be expressed in terms of the LL
parameter $K$.   
We proposed a setup which simultaneously allows to measure $n(V_g)$,
and thus $w(\Gamma)$, as well as the temperature dependence of the
linear conductance $G(T)$ through the QD. Off-resonance the latter 
is also governed by power-law scaling with an exponent which can be
expressed in terms of $K$. Consistency of the extracted $K$'s  
would provide strong evidence for the experimental observation of LL
physics.

\section*{Acknowledgments}
This work was supported by the Deutsche Forschungsgemeinschaft (SFB
602).


\begin{thebibliography}{*}
\bibitem{footnoteresowidth} For asymmetric barriers $t_l^2 \neq
  t_r^2$, $0<K<1$,  and in the infrared limit $T \to 0$ as well 
as $\delta \to 0$, the linear conductance through a double barrier 
vanishes for all $V_g$. This holds even at $V_g=0$ where for $K=1$ a 
resonance peak of transmission $4 t_l^2 t_r^2/(t_l^2+t_r^2)$ is 
located. At finite $T$ or $\delta$ a transmssion peak of 
nonzero height remains at $V_g=0$. For symmetric barriers 
$t^2_l=t^2_r$ a resonance of unitary transmission is located
at $V_g=0$, but the resonance width $\sigma$ vanishes $\sim \,
\mbox{max}\{ T, \delta\}^{1-K}$.\cite{doublebarrier}
\bibitem{footnotewdef} We here (for the fRG data) defined the
  width $w$ of $n(V_g)$ as the full width at half maximum of
  $-dn(V_g)/dV_g$. Concerning the scaling properties of $w(\Gamma)$
  this definition is equivalent to the one used in the derivation of
  Eq.~(\ref{wid}).
\end{thebibliography}
\end{document}